%% file: URSI_Summary_Paper_Template.tex
\documentclass[summary]{ursi}
\input{include.tex}
\input{glossary.tex}

\input{colormaps_new.tex}
\title{Near field Exposure Assessment of Complex Anatomical Structures in 5G Bands}

\author{G. Sacco*\affref{ref1}, A. Kapetanovi\'{c}\affref{ref1}, D. Poljak\affref{ref2}
  and M. Zhadobov\affref{ref1}}

\affiliation{%
  \aff{ref1}{Institut d’Électronique et des Technologies du numéRique (IETR), University of Rennes 1 CNRS, UMR 6164, F-35000 Rennes, France; e-mail: giulia.sacco@univ-rennes.fr; maxim.zhadobov@univ-rennes.fr}
  \aff{ref2}{Faculty of Electrical Engineering, Mechanical Engineering and Naval Architecture (FESB), Univeristy of Split, 21000 Split, Croatia; e-mail: akapet00@gmail.com; dpoljak@fesb.hr}
}

\begin{document}

\maketitle

\begin{abstract}
  With the proliferation of 5G wireless networks, the population is increasingly exposed to frequencies approaching the \gls{mmw} range. Human ears are among the most exposed body parts. This paper proposes an analysis of the ear exposure in the near field using an anatomical model in presence of different \gls{em} sources (vertical dipole, horizontal dipole, $4\times4$ array of vertical dipoles, and $4\times4$ array of horizontal dipoles). This study demonstrates that, for a given input power and antenna-ear distance, the absorbed power density ($S_{\text{ab}}$) induced by a dipole antenna array is up to $3.9$ times higher than the one produced by a single dipole. $S_{\text{ab}}$ is only slightly sensitive to the dipole orientation (vertical or horizontal) resulting in relatively weak variations (up to \SI{7}{\percent}).
\end{abstract}
\glsresetall
\section{Introduction}
With the proliferation of 5G wireless networks, new frequency bands (including \SIrange[range-units=single,range-phrase=--]{24}{28}{\GHz} range) have been proposed to enable reduced latency and higher data rates. This will introduce to the environmental \gls{em} spectrum frequencies to which population has never been exposed so far.

For exposure assessment at \gls{mmw}, due to a high computational cost and the shallow penetration depth (e.g., roughly \SI{0.85}{\mm} at \SI{30}{\GHz}), mainly mono- or multi-layer planar tissue models have been used \cite{sasakiMonteCarloSimulations2017, christRFinducedTemperatureIncrease2020}. However, some of the most exposed body parts cannot be accurately modelled as planar \cite{saccoExposureLevelsInduced2022}. In \cite{colellaNumericalComparisonPlane2022}, the authors compared the exposure of anatomical models of abdomen and wrist at \SI{24}{\GHz} with multi-layer tissue models. The findings demonstrated that the electric field distribution could not be accurately reproduced by a planar model. In \cite{diaoAssessmentAbsorbedPower2020}, the exposure of forearm was investigated in the \SIrange[range-phrase=--,range-units=single]{6}{60}{\GHz} range. The \gls{em} energy absorption in the head and hand at \SI{60}{\GHz} was discussed in \cite{guraliucNearfieldUserExposure2017}. The exposure of body parts with smaller curvature radii, comparable to the wavelength in the \gls{mmw} range, was performed in \cite{saccoExposureLevelsInduced2022} with simplified cylindrical based models and in \cite{kapetanovicAreaAveragedTransmittedAbsorbed2022} for an anatomical ear model, considering a plane wave as source. 

This paper aims at quantifying the power absorption in the human ear in the near field for different realistic antenna sources.%
\begin{figure*}
	\centering
	\subfloat[]{\label{fig:vertical_dipole}\resizebox{0.485\columnwidth}{!}{\input{figures/vertical_dipole.tex}}}
	\subfloat[]{\label{fig:horizontal_dipole}\resizebox{0.4\columnwidth}{!}{\input{figures/horizontal_dipole.tex}}}
	\subfloat[]{\label{fig:vertical_array}\resizebox{0.4\columnwidth}{!}{\input{figures/vertical_array.tex}}}
	\subfloat[]{\label{fig:horizontal_array}\resizebox{0.4\columnwidth}{!}{\input{figures/horizontal_array.tex}}}
	\caption{Exposure scenarii: (\protect\subref{fig:vertical_dipole}) vertical dipole, (\protect\subref{fig:horizontal_dipole}) horizontal dipole, (\protect\subref{fig:vertical_array}) $4\times 4$ array of vertical dipoles, and (\protect\subref{fig:horizontal_array}) $4\times 4$ array of horizontal dipoles.}
	\label{fig:geometry}
\end{figure*}
\begin{figure*}
	\centering
	\subfloat[]{\label{fig:DipoleVertical_d5mm_labels}\resizebox{0.65\columnwidth}{!}{\includegraphics{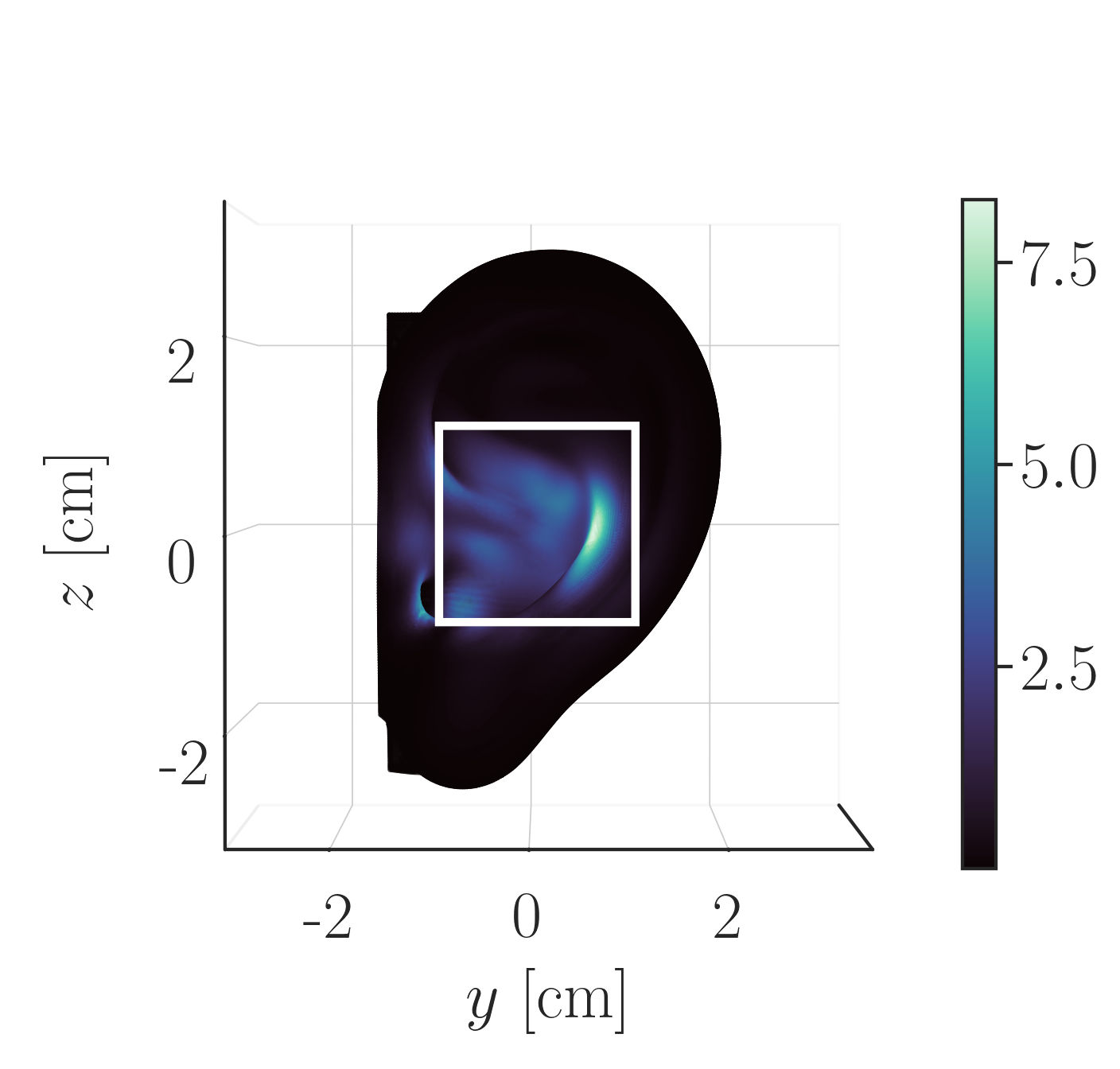}}}
	\subfloat[]{\label{fig:DipoleVertical_d10mm_labels}\resizebox{0.65\columnwidth}{!}{\includegraphics{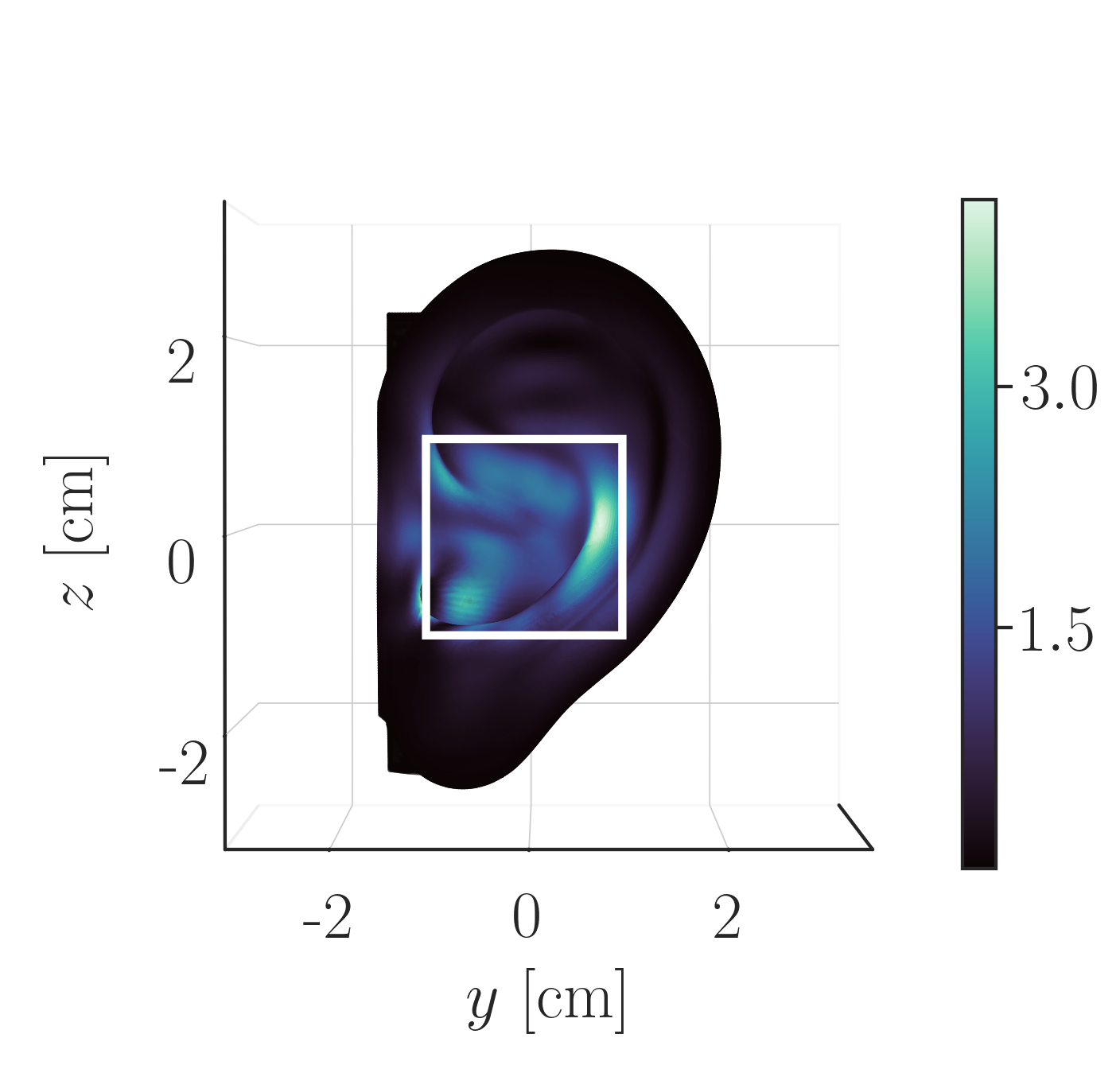}}}
	\subfloat[]{\label{fig:DipoleVertical_d15mm_labels}\resizebox{0.725\columnwidth}{!}{\includegraphics{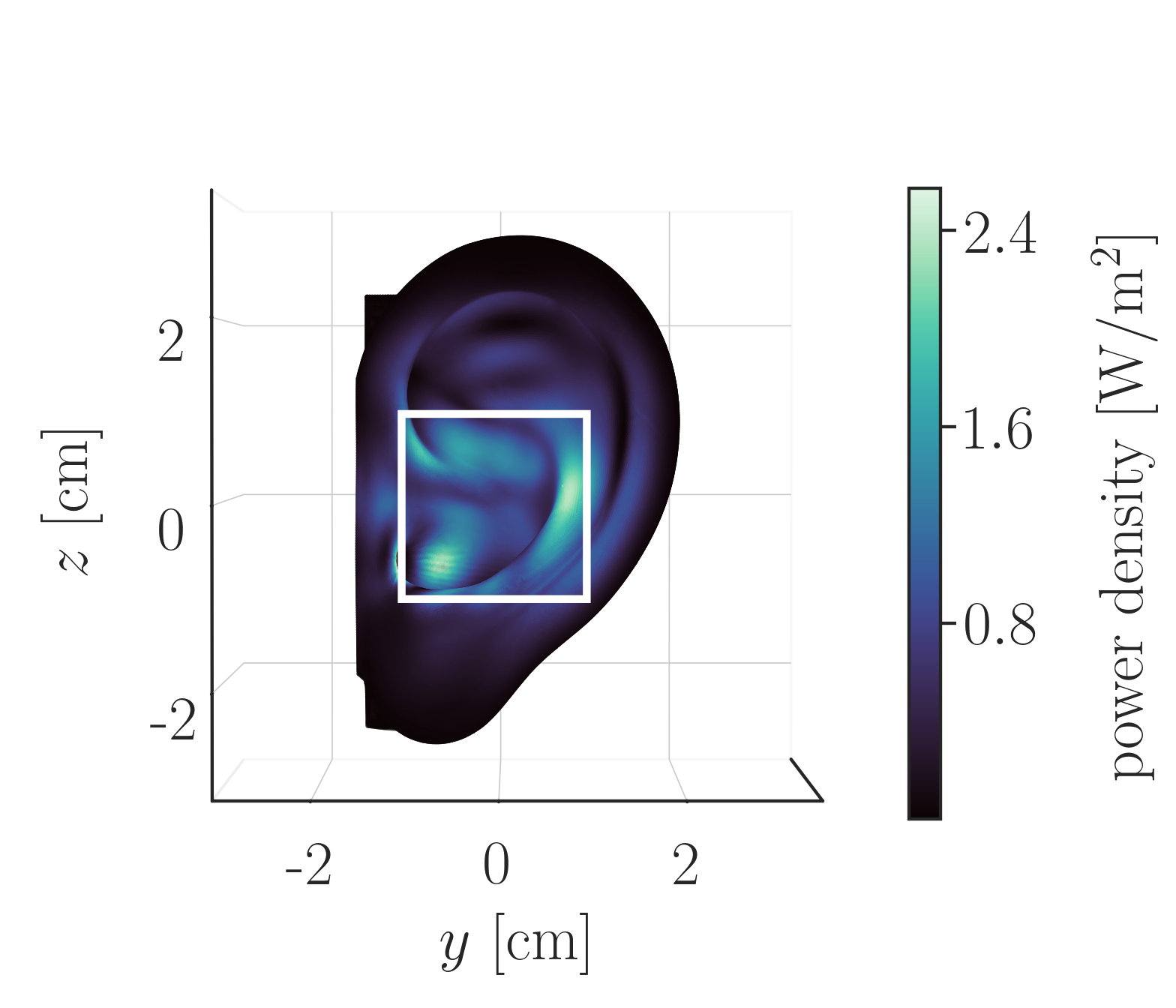}}}\\
	\subfloat[]{\label{fig:ArrayVertical_d5mm_labels}\resizebox{0.65\columnwidth}{!}{\includegraphics{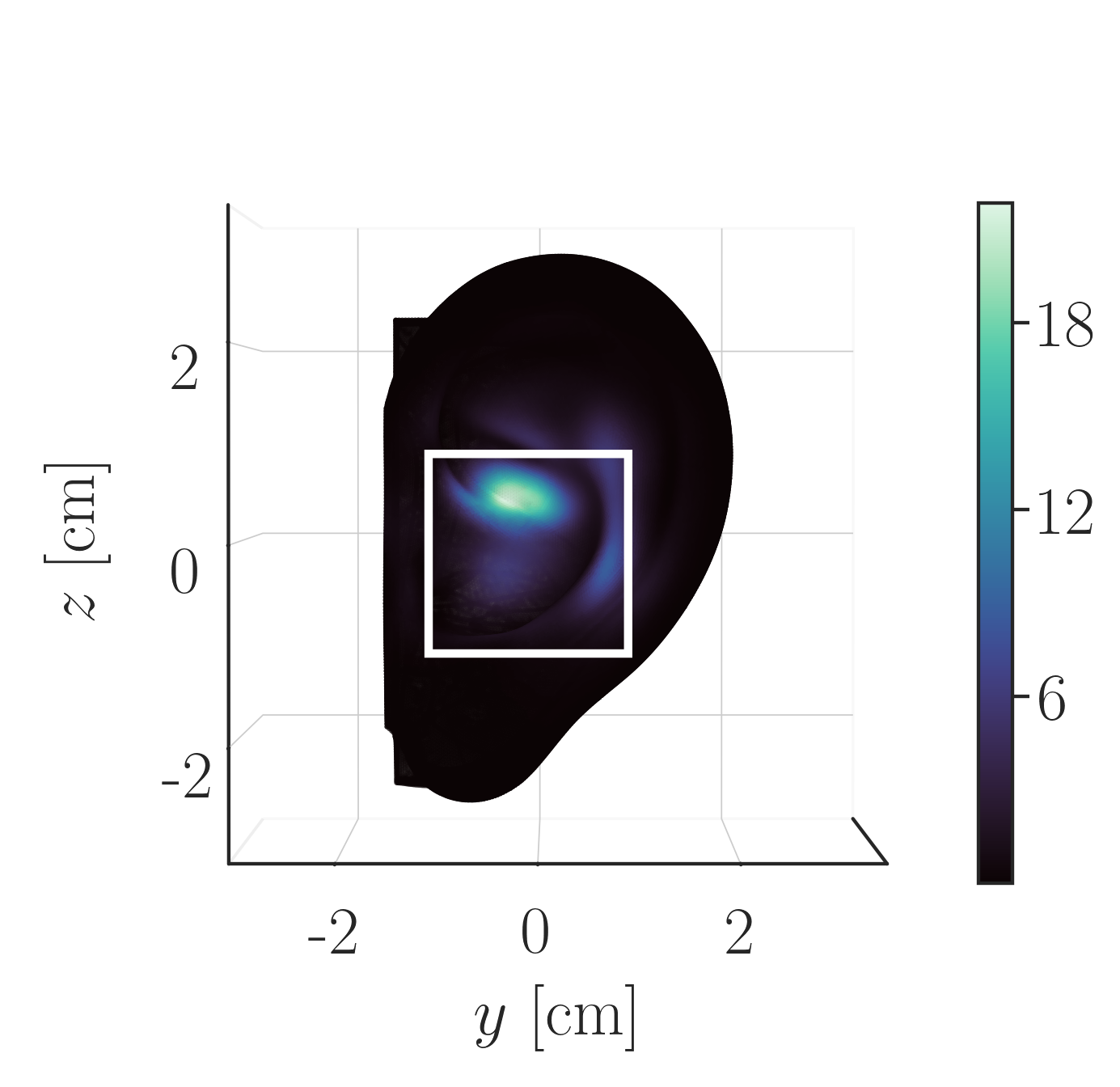}}}
	\subfloat[]{\label{fig:ArrayVertical_d10mm_labels}\resizebox{0.65\columnwidth}{!}{\includegraphics{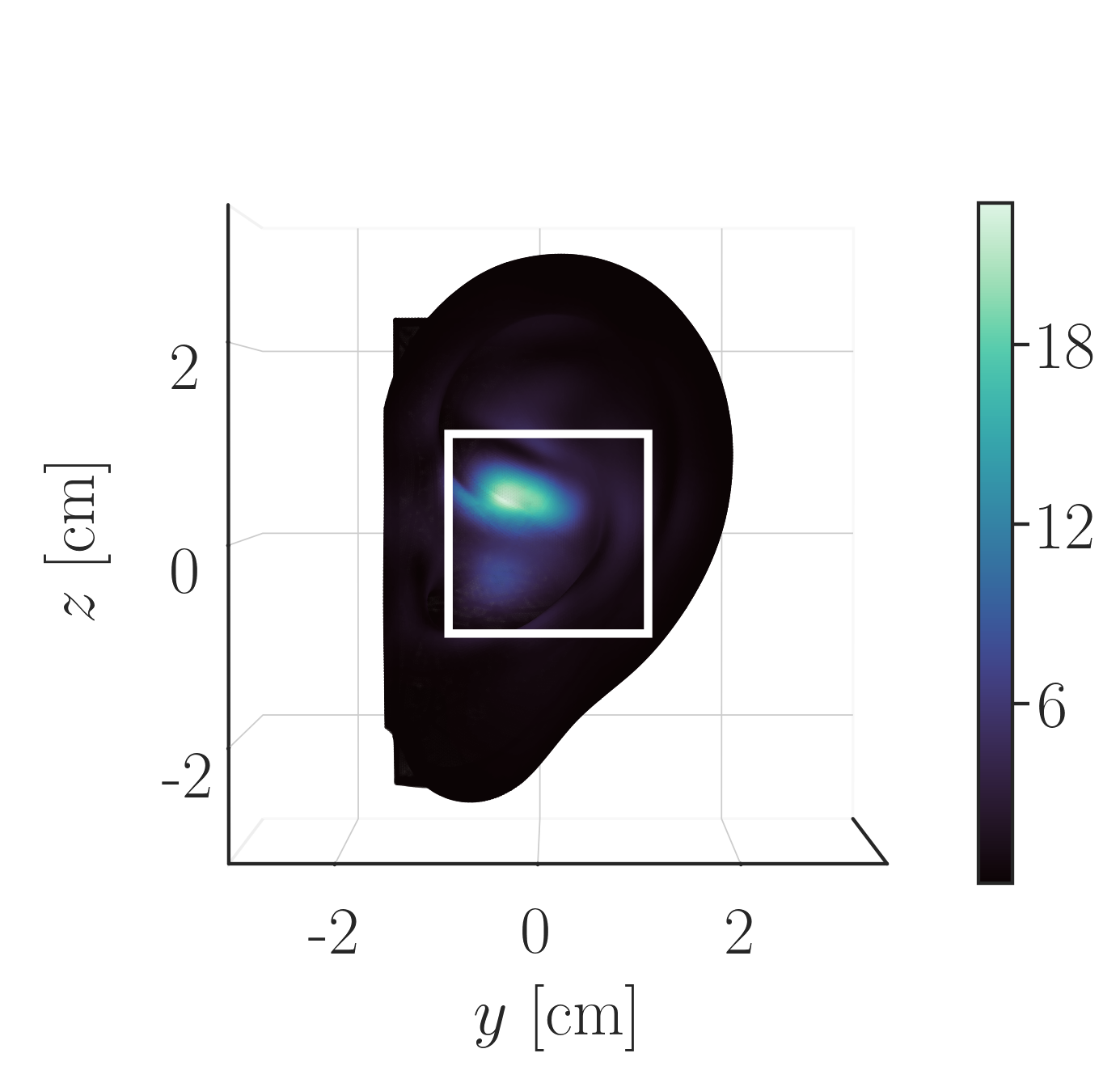}}}
	\subfloat[]{\label{fig:ArrayVertical_d15mm_labels}\resizebox{0.725\columnwidth}{!}{\includegraphics{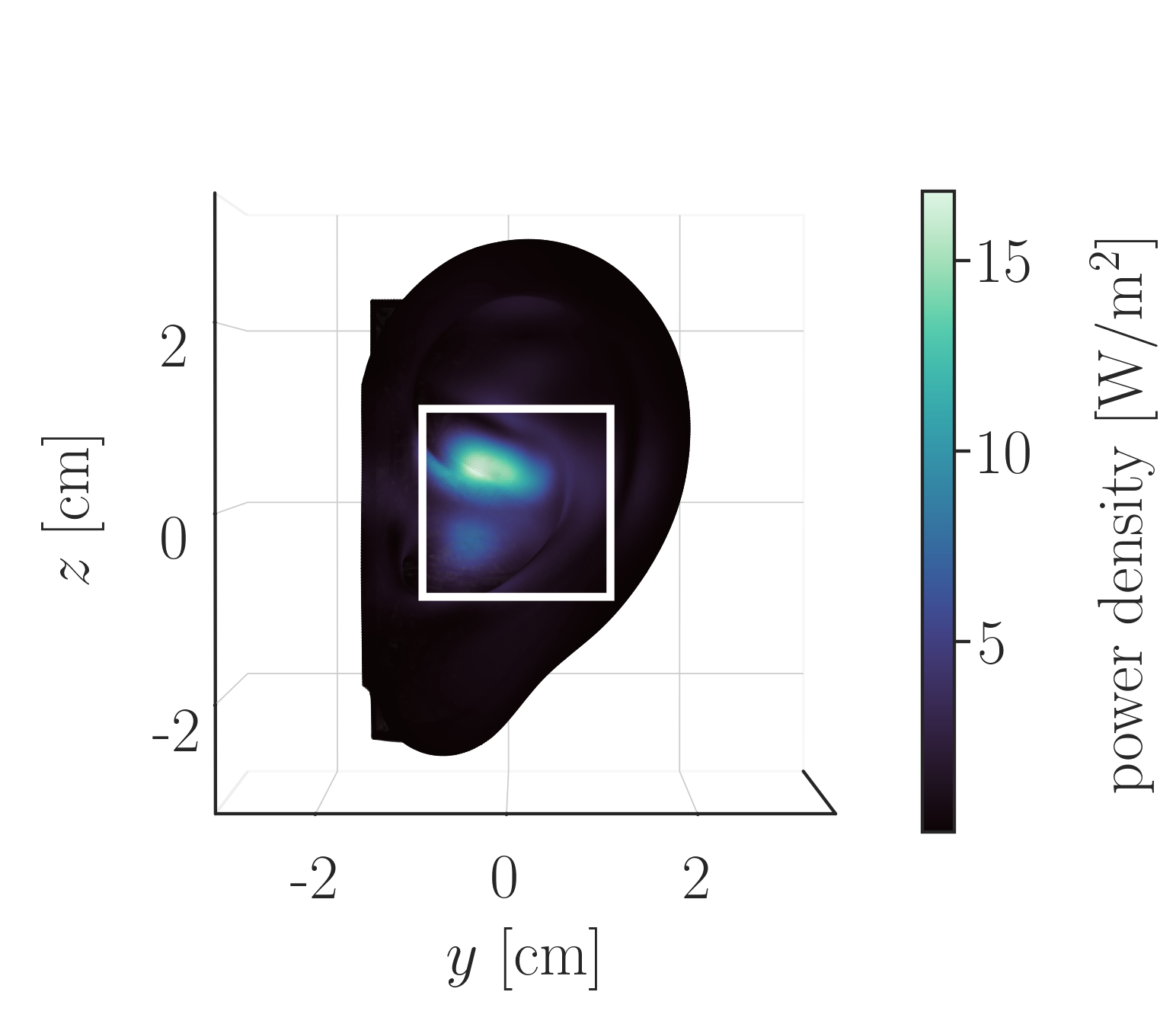}}}
	\caption{$S_\text{ab}$ distribution on the ear surface in presence of a vertical dipole placed at (\protect\subref{fig:DipoleVertical_d5mm_labels}) \SI{5}{\mm}, (\protect\subref{fig:DipoleVertical_d10mm_labels}) \SI{10}{\mm}, and (\protect\subref{fig:DipoleVertical_d15mm_labels}) \SI{15}{\mm} from the model surface and of a $4\times4$ array of vertical dipoles placed at (\protect\subref{fig:ArrayVertical_d5mm_labels}) \SI{5}{\mm}, (\protect\subref{fig:ArrayVertical_d10mm_labels}) \SI{10}{\mm}, and (\protect\subref{fig:ArrayVertical_d15mm_labels}) \SI{15}{\mm} from the model surface.}
	\label{fig:APD_distribution}
\end{figure*}
\section{Materials and Methods}
\subsection{EM Model and Scenario}
We considered a homogenous anatomical model of the adult ear with the typical dimensions and the complex permittivity of dry skin at \SI{26}{\GHz} ($17.71-j16.87$) \cite{gabrielCompilationDielectricProperties1996}. 

To evaluate the \gls{em} power deposition, we used $4$ radiating sources: vertical dipole, horizontal dipole, $4\times4$ array of vertical dipoles, and $4\times4$ array of horizontal dipoles (\Cref{fig:geometry}). 

The input power was set to \SI{10}{\mW}. The antenna sources are centred with respect to the ear model in the $yz$ plane and the distance from the model surface ($x$ direction) was varied from \SI{5}{\mm} to \SI{15}{\mm} with a \SI{5}{\mm} step. The model was simulated using COMSOL Multiphysics with the \gls{fem} technique. It was discretized with a tetrahedral mesh with a maximum cell size of $\lambda/8$. \Gls{pml} was used as boundary condition.

\subsection{Averaged Absorbed Power}
Above \SI{6}{\GHz}, according to the international guidelines of \gls{icnirp} and \gls{ieee}, the main dosimetric quantity is the absorbed power density \cite{internationalcommissiononnon-ionizingradiationprotectionicnirpGuidelinesLimitingExposure2020,internationalelectricalandelectronicsengineersieeeIEEEStandardSafety2019}
\begin{equation}
	\centering
	S_\text{ab} = \frac{1}{2A} \iint_{A} \Re \big[\vt{E}(y, z) \times \vt{H}^*(y, z) \big] \; \boldsymbol{\hat n} \; \mathrm{d}A\,,
	\label{eqn:apd_2} 
\end{equation}
where $\vt{E}$ and $\vt{H}$ are the peak values of the electric and magnetic fields on the model surface, respectively, $\Re$ is the real part operator, $*$ is the complex conjugate operator, $A$ is the averaging area, and $\boldsymbol{\hat n} \; \mathrm{d}A$ is the integral variable vector, $\boldsymbol{\hat n}$ being the unit vector field normal to the surface.

$S_\text{ab}$ was evaluated over the entire ear surface to find the coordinates of the most exposed area (worst-case scenario). The averaging area used to compute $S_{\text{ab}}$ is the area conformal to the ear surface and limited by a $2\times\SI{2}{\cm}$ square, that is typically larger than \SI{4}{\cm\squared} due to the non planarity of the model \cite{kapetanovicAreaAveragedTransmittedAbsorbed2022}.

\section{Results}
 The $S_\text{ab}$ distribution over the ear surface for a vertical dipole and a $4\times4$ array of vertical dipoles placed at \SI{5}{\mm}, \SI{10}{\mm}, and \SI{15}{\mm} from the model surface is reported in \Cref{fig:APD_distribution}. Similar distributions were obtained also for a horizontal dipole and dipole array (not shown in \Cref{fig:APD_distribution}). The white squares highlight the considered averaging area, which corresponds to the highest $S_\text{ab}$ (worst-case). As it is possible to notice, the dipole array produces stronger $S_\text{ab}$ values and the most exposed area is concentrated in a few millimetres region.

\begin{figure}
  \centering
  \resizebox{0.45\textwidth}{!}{\input{data/APD_dipole.tex}}
  \caption{$S_\text{ab}$ as a function of the distance.}
  \label{fig:APD_dipole}
\end{figure}
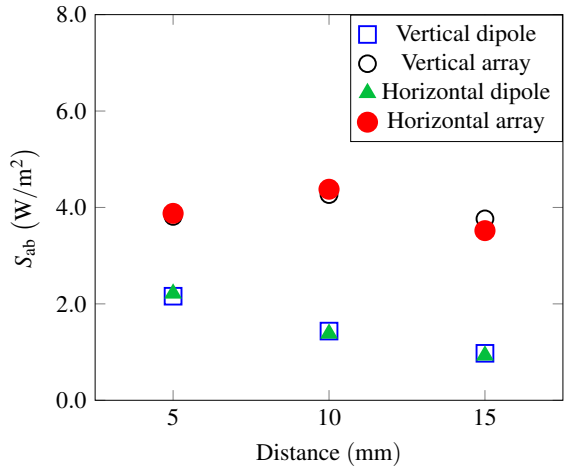

 The results of $S_\text{ab}$ averaged on the white squares as a function of the distance is shown in \Cref{fig:APD_dipole}.
For a single dipole, $S_\text{ab}$ decreases monotonically with the distance. However, when considering a $4\times 4$ array, regardless of the dipoles' orientation, $S_\text{ab}$ is greater at the separation distance of \SI{10}{\mm} comparing to \SI{5}{\mm}. This may be explained by the different antenna-ear interactions occurring in the near field \cite{zianeAntennaBodyCoupling2020}.
For a given distance, the difference in exposure between the two antenna orientations is limited to \SI{4.7}{\percent} and \SI{7}{\percent} for the single dipole and dipole array, respectively. For the same input power, $S_\text{ab}$ is up to $3.9$ times higher for the array than for the single dipole for vertical orientation and $3.8$ times for horizontal orientation.
\section{Conclusion}
This paper evaluates the near field exposure of the ear anatomical model at \SI{26}{\GHz}. A dipole and a $4\times 4$ array of dipoles with vertical and horizontal orientations were considered as \gls{em} sources.

For the same input power (\SI{10}{\mW}) and distance, the array was responsible for higher values of $S_{\text{ab}}$ (up to \SI{4.4}{\W\per\square\m}) than the single dipole (up to \SI{2.1}{\W\per\square\m}), while the difference in terms of antenna orientation (vertical or horizontal) had almost no impact on $S_{\text{ab}}$ (up to \SI{7}{\percent} difference). The quantification of the resulting temperature rise is out of the scope of this paper, but constitutes one of its perspectives.
\section*{Acknowledgements}

This research was supported by the European Regional Development Fund under the grant KK.01.1.1.01.0009 (DATACROSS), the French National Research Program for Environmental and Occupational Health of ANSES under Grant 2018/2 RF/07 through the NEAR 5G Project, and the European Union’s Horizon Europe research and innovation program through the Marie Sk\l odowska-Curie IN-SIGHT project N$^\circ$101063966.

\bibliographystyle{IEEEtran}
\bibliography{bib}

\end{document}

%% file: include.tex
\usepackage{amsmath}
\usepackage{amssymb}
\usepackage{array}
\usepackage[english]{babel}
\usepackage{blindtext}
\usepackage{bm}
\usepackage{booktabs}
\usepackage{caption}
\usepackage{circuitikz}
\usepackage{cite}
\usepackage{comment}
\usepackage{enumitem}
\usepackage[T1]{fontenc}
\usepackage{framed}
\usepackage[acronym]{glossaries-prefix}
\usepackage{graphicx}
\usepackage{hyperref}
\hypersetup{
	colorlinks=true,
	linkcolor=black,
	filecolor=magenta,      
	urlcolor=mediumblue,
	citecolor=black,
	pdftitle={Overleaf Example},
	pdfpagemode=FullScreen,
}
\usepackage[capitalise]{cleveref}
\usepackage[utf8]{inputenc}
\usepackage{multirow}
\usepackage{pgf}
\usepackage{pgfplots}
\usepackage{pgfplotstable}
\usepackage{pst-plot}
\usepackage{pst-3dplot}
\usepackage{relsize}
\usepackage[detect-all,separate-uncertainty = true,multi-part-units=single]{siunitx}
\usepackage{soul}
\usepackage{steinmetz}
\usepackage{subcaption}
\usepackage{times}
\usepackage{tikz}
\usepackage{tikz-3dplot}
\usepackage{titlesec}
\usepackage{todonotes}
\usepackage{url}
\usepackage{xparse}
\usepackage[]{xcolor,colortbl}
\usepackage{wrapfig}

\newcommand{\vt}[1]{\bm{#1}}

\usetikzlibrary{external}
\input{colormaps_new.tex}

\usetikzlibrary{arrows}
\usetikzlibrary{arrows.meta}
\usetikzlibrary{calc}
\pgfplotsset{compat=newest}
\usetikzlibrary{decorations.markings}
\usepgfplotslibrary{external}
\usetikzlibrary{fadings}
\usepgfplotslibrary{fillbetween}
\usepgfplotslibrary{patchplots}
\usetikzlibrary{patterns}
\usepgfplotslibrary{polar}
\usetikzlibrary{positioning}
\usetikzlibrary{spy}
\usepgfplotslibrary{units}

\DeclareSIUnit[number-unit-product = ]\percent{\char`\%}
\DeclareSIUnit{\dBi}{dBi}
\DeclareSIUnit{\dBm}{dBm}
\DeclareSIUnit{\year}{years}
\DeclareSIUnit{\bpm}{BPM}

\definecolor{mediumblue}{rgb}{0.0, 0.0, 0.8}
\definecolor{darkred}{rgb}{0.55, 0.0, 0.0}

%% file: colormaps_new.tex
\pgfplotsset{
  unbounded coords=discard,
  colormap/magma/.style={%
    /pgfplots/colormap={magma}{%
      rgb=(0.001462, 0.000466, 0.013866)
      rgb=(0.035520, 0.028397, 0.125209)
      rgb=(0.102815, 0.063010, 0.257854)
      rgb=(0.191460, 0.064818, 0.396152)
      rgb=(0.291366, 0.064553, 0.475462)
      rgb=(0.384299, 0.097855, 0.501002)
      rgb=(0.475780, 0.134577, 0.507921)
      rgb=(0.569172, 0.167454, 0.504105)
      rgb=(0.664915, 0.198075, 0.488836)
      rgb=(0.761077, 0.231214, 0.460162)
      rgb=(0.852126, 0.276106, 0.418573)
      rgb=(0.925937, 0.346844, 0.374959)
      rgb=(0.969680, 0.446936, 0.360311)
      rgb=(0.989363, 0.557873, 0.391671)
      rgb=(0.996580, 0.668256, 0.456192)
      rgb=(0.996727, 0.776795, 0.541039)
      rgb=(0.992440, 0.884330, 0.640099)
      rgb=(0.987053, 0.991438, 0.749504)
    },
  },
  colormap/inferno/.style={%
    /pgfplots/colormap={inferno}{%
      rgb=(0.001462, 0.000466, 0.013866)
      rgb=(0.037668, 0.025921, 0.132232)
      rgb=(0.116656, 0.047574, 0.272321)
      rgb=(0.217949, 0.036615, 0.383522)
      rgb=(0.316282, 0.053490, 0.425116)
      rgb=(0.410113, 0.087896, 0.433098)
      rgb=(0.503493, 0.121575, 0.423356)
      rgb=(0.596940, 0.154848, 0.398125)
      rgb=(0.688653, 0.192239, 0.357603)
      rgb=(0.775059, 0.239667, 0.303526)
      rgb=(0.851384, 0.302260, 0.239636)
      rgb=(0.912966, 0.381636, 0.169755)
      rgb=(0.956852, 0.475356, 0.094695)
      rgb=(0.981895, 0.579392, 0.026250)
      rgb=(0.987464, 0.690366, 0.079990)
      rgb=(0.973088, 0.805409, 0.216877)
      rgb=(0.947594, 0.917399, 0.410665)
      rgb=(0.988362, 0.998364, 0.644924)
    },
  },
  colormap/plasma/.style={%
    /pgfplots/colormap={plasma}{%
      rgb=(0.050383, 0.029803, 0.527975)
      rgb=(0.186213, 0.018803, 0.587228)
      rgb=(0.287076, 0.010855, 0.627295)
      rgb=(0.381047, 0.001814, 0.653068)
      rgb=(0.471457, 0.005678, 0.659897)
      rgb=(0.557243, 0.047331, 0.643443)
      rgb=(0.636008, 0.112092, 0.605205)
      rgb=(0.706178, 0.178437, 0.553657)
      rgb=(0.768090, 0.244817, 0.498465)
      rgb=(0.823132, 0.311261, 0.444806)
      rgb=(0.872303, 0.378774, 0.393355)
      rgb=(0.915471, 0.448807, 0.342890)
      rgb=(0.951344, 0.522850, 0.292275)
      rgb=(0.977856, 0.602051, 0.241387)
      rgb=(0.992541, 0.687030, 0.192170)
      rgb=(0.992505, 0.777967, 0.152855)
      rgb=(0.974443, 0.874622, 0.144061)
      rgb=(0.940015, 0.975158, 0.131326)
    },
  },
  colormap/jpeBlWhBl/.style={%
    /pgfplots/colormap={jpeReWhBl}{%
      rgb=(0.094957, 0.211745, 0.949475)
      rgb=(0.149676, 0.288182, 0.889340)
      rgb=(0.210373, 0.340936, 0.846637)
      rgb=(0.268241, 0.383971, 0.815988)
      rgb=(0.322709, 0.422216, 0.794156)
      rgb=(0.374335, 0.457967, 0.779078)
      rgb=(0.423724, 0.492464, 0.769388)
      rgb=(0.471369, 0.526452, 0.764155)
      rgb=(0.517653, 0.560415, 0.762732)
      rgb=(0.562866, 0.594688, 0.764671)
      rgb=(0.607234, 0.629520, 0.769662)
      rgb=(0.650934, 0.665108, 0.777508)
      rgb=(0.694107, 0.701609, 0.788103)
      rgb=(0.736875, 0.739156, 0.801431)
      rgb=(0.779363, 0.777851, 0.817590)
      rgb=(0.821732, 0.817752, 0.836891)
      rgb=(0.863910, 0.858922, 0.860421)
      rgb=(0.902313, 0.903243, 0.885111)
      rgb=(0.879049, 0.855058, 0.839087)
      rgb=(0.860371, 0.806330, 0.794603)
      rgb=(0.843643, 0.758010, 0.750027)
      rgb=(0.828754, 0.709912, 0.704933)
      rgb=(0.815142, 0.662021, 0.659169)
      rgb=(0.802302, 0.614305, 0.612550)
      rgb=(0.789774, 0.566724, 0.564903)
      rgb=(0.777122, 0.519227, 0.516091)
      rgb=(0.763930, 0.471751, 0.466024)
      rgb=(0.749803, 0.424208, 0.414668)
      rgb=(0.734369, 0.376470, 0.362052)
      rgb=(0.717286, 0.328322, 0.308250)
      rgb=(0.698243, 0.279394, 0.253349)
      rgb=(0.676967, 0.228987, 0.197339)
      rgb=(0.653216, 0.175627, 0.139788)
      rgb=(0.626782, 0.115406, 0.078551)
      rgb=(0.597480, 0.029268, 0.008874)
    },
  },
  colormap/jpeWhRe/.style={%
    /pgfplots/colormap={jpeWhRe}{%
      rgb=(0.902313, 0.903243, 0.885111)
      rgb=(0.879049, 0.855058, 0.839087)
      rgb=(0.860371, 0.806330, 0.794603)
      rgb=(0.843643, 0.758010, 0.750027)
      rgb=(0.828754, 0.709912, 0.704933)
      rgb=(0.815142, 0.662021, 0.659169)
      rgb=(0.802302, 0.614305, 0.612550)
      rgb=(0.789774, 0.566724, 0.564903)
      rgb=(0.777122, 0.519227, 0.516091)
      rgb=(0.763930, 0.471751, 0.466024)
      rgb=(0.749803, 0.424208, 0.414668)
      rgb=(0.734369, 0.376470, 0.362052)
      rgb=(0.717286, 0.328322, 0.308250)
      rgb=(0.698243, 0.279394, 0.253349)
      rgb=(0.676967, 0.228987, 0.197339)
      rgb=(0.653216, 0.175627, 0.139788)
      rgb=(0.626782, 0.115406, 0.078551)
      rgb=(0.597480, 0.029268, 0.008874)
    },
  },
	colormap/viridis_new/.style={%
	/pgfplots/colormap={viridis_new}{%
	rgb=(0.267,0.00487,0.32942)
	rgb=(0.28192,0.08966,0.41241)
	rgb=(0.28026,0.1657,0.4765)
	rgb=(0.26366,0.23763,0.51877)
	rgb=(0.23744,0.3052,0.54192)
	rgb=(0.20862,0.36775,0.55267)
	rgb=(0.18225,0.42618,0.55711)
	rgb=(0.1592,0.48224,0.55807)
	rgb=(0.13777,0.53749,0.5549)
	rgb=(0.12115,0.59274,0.54465)
	rgb=(0.12808,0.64775,0.5235)
	rgb=(0.18065,0.7014,0.48819)
	rgb=(0.27415,0.75198,0.4366)
	rgb=(0.39517,0.79747,0.36775)
	rgb=(0.53561,0.83578,0.2819)
	rgb=(0.68895,0.86545,0.18272)
	rgb=(0.84557,0.88733,0.0997)
	rgb=(0.99324,0.90616,0.14394)
	              		rgb=(1.0, 1.0, 1.0)
		},
},
  colormap/viridis,
}


%% file: glossary.tex
\newacronym{5g}{5G}{5\textsuperscript{th} generation}
\newacronym{6g}{6G}{6\textsuperscript{th} generation}
\newacronym{aal}{AAL}{ambient-assisted living}
\newacronym{ban}{BAN}{body area network}
\newacronym{bpm}{BPM}{beats per minute}
\newacronym{cs}{CS}{compressed sensing}
\newacronym{cw}{CW}{continuous wave}
\newacronym{dnm}{DNM}{distributed network method}
\newacronym{ebg}{EBG}{electromagnetic band-gap}
\newacronym[prefixfirst={a\ },
prefix={an\ }]{em}{EM}{electromagnetic}
\newacronym{eta}{ETA}{electronic travel aid}
\newacronym{fdtd}{FDTD}{finite difference time domain}
\newacronym{fem}{FEM}{finite element method}
\newacronym{fit}{FIT}{finite integration technique}
\newacronym[prefixfirst={a\ },
prefix={an\ }]{fmcw}{FMCW}{frequency-modulated continuous wave}
\newacronym{hpc}{HPC}{high-performance computing}
\newacronym{icrp}{ICRP}{Commission on Radiological Protection}
\newacronym{icnirp}{ICNIRP}{International Commission on Non-ionizing Radiation Protection}
\newacronym{ieee}{IEEE}{International Electrical and Electronics Engineers}
\newacronym{imu}{IMU}{inertial measurement unit}
\newacronym{ic}{IC}{integrated circuit}
\newacronym{iot}{IoT}{internet of things}
\newacronym{ipd}{IPD}{incident power density}
\newacronym{itu}{ITU}{International Telecommunication Union}
\newacronym{mtnm}{MTNM}{meshed transport network method}
\newacronym{mimo}{MIMO}{multiple-input and multiple-output}
\newacronym{mlfmm}{MLFMM}{multipole fast multipole method}
\newacronym{mmw}{mmW}{millimeter-wave}
\newacronym{mom}{MoM}{method of moments}
\newacronym{nn}{NN}{neural network}
\newacronym{pdms}{PDMS}{polydimethylsiloxane}
\newacronym{pi}{PI}{principal investigator}
\newacronym{pml}{PML}{perfectly matched layer}
\newacronym{ppg}{PPG}{photoplethysmograph}
\newacronym{rbm}{RBM}{random body motion}
\newacronym{rsm}{RSM}{radar selfmotion}
\newacronym{sar}{SAR}{specific absorption rate}
\newacronym{sc}{SC}{stratum corneum}
\newacronym{siso}{SISO}{single-input and single-output}
\newacronym{sll}{SLL}{side lobe level}
\newacronym{smd}{SMD}{surface mounting device}
\newacronym{snr}{SNR}{signal-to-noise ratio}
\newacronym{ro}{RO}{research objective}
\newacronym{te}{TE}{transvers electric}
\newacronym{tm}{TM}{transvers magnetic}
\newacronym{tbw}{TBW}{total body water}
\newacronym{uwb}{UWB}{ultra-wideband}
\newacronym{wp}{WP}{work package}
\newacronym{mae}{MAE}{mean absolute error}

%% file: figures/vertical_dipole.tex
\definecolor{aurometalsaurus}{rgb}{0.43, 0.5, 0.5}
\definecolor{pastelpink}{rgb}{1.0,  0.8353,  0.8353}

\begin{tikzpicture}
	\node[anchor=center] at (0,0) {\includegraphics[scale=0.3]{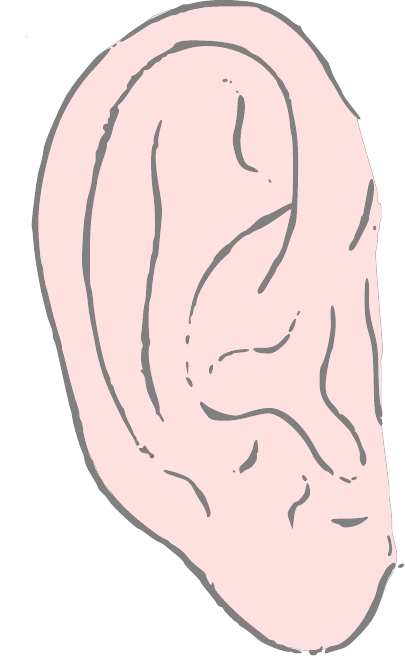}};
	\draw[blue,line width=0.35mm](0,-0.15)--(0,-0.01);
	\draw[line width=0.35mm, blue](0,0.01)--(0,0.15);
	
	\draw[-latex, black,line width=0.3mm](-1.3,-1.5)--(-0.7,-1.5);
	\draw[-latex, black,line width=0.3mm](-1.3,-1.5)--(-1.3,-0.9);
	\draw[line width=0.3mm, black] (-1.3,-1.5) circle (0.15);
	\fill[line width=0.3mm, black] (-1.3,-1.5)  circle (0.075);
	\node at (-0.65,-1.7){\textbf{y}};
	\node at (-1.5,-1){\textbf{z}};
	\node at (-1.6,-1.7){\textbf{x}};
\end{tikzpicture}

%% file: figures/horizontal_dipole.tex
\definecolor{aurometalsaurus}{rgb}{0.43, 0.5, 0.5}
\definecolor{pastelpink}{rgb}{1.0,  0.8353,  0.8353}

\begin{tikzpicture}
	\node[anchor=center] at (3,0) {\includegraphics[scale=0.3]{figures/ear_black.pdf}};
	\draw[line width=0.35mm, red](2.85,0)--(2.99,0);
	\draw[line width=0.35mm, red](3.01,0)--(3.15,0);
		
\end{tikzpicture}

%% file: figures/vertical_array.tex
\definecolor{aurometalsaurus}{rgb}{0.43, 0.5, 0.5}
\definecolor{pastelpink}{rgb}{1.0,  0.8353,  0.8353}

\begin{tikzpicture}
	\node[anchor=center] at (6,0) {\includegraphics[scale=0.3]{figures/ear_black.pdf}};
	
	\draw[blue,line width=0.35mm](5.85,0.56)--(5.85,0.42);
	\draw[blue,line width=0.35mm](5.85,0.58)--(5.85,0.72);
	\draw[line width=0.35mm, blue](6.15,0.56)--(6.15,0.42);
	\draw[line width=0.35mm, blue](6.15,0.58)--(6.15,0.72);
	\draw[line width=0.35mm, blue](6.45,0.56)--(6.45,0.42);
	\draw[line width=0.35mm, blue](6.45,0.58)--(6.45,0.72);
	\draw[line width=0.35mm, blue](5.55,0.56)--(5.55,0.42);
	\draw[line width=0.35mm, blue](5.55,0.58)--(5.55,0.72);
	
	\draw[line width=0.35mm, blue](5.85,0.04)--(5.85,0.18);
	\draw[line width=0.35mm, blue](5.85,0.2)--(5.85,0.34);
	\draw[line width=0.35mm, blue](6.15,0.04)--(6.15,0.18);
	\draw[line width=0.35mm, blue](6.15,0.2)--(6.15,0.34);
	\draw[line width=0.35mm, blue](6.45,0.04)--(6.45,0.18);
	\draw[line width=0.35mm, blue](6.45,0.2)--(6.45,0.34);
	\draw[line width=0.35mm, blue](5.55,0.04)--(5.55,0.18);
	\draw[line width=0.35mm, blue](5.55,0.2)--(5.55,0.34);
	
	\draw[line width=0.35mm, blue](5.85,-0.34)--(5.85,-0.2);
	\draw[line width=0.35mm, blue](5.85,-0.18)--(5.85,-0.04);
	\draw[line width=0.35mm, blue](6.15,-0.34)--(6.15,-0.2);
	\draw[line width=0.35mm, blue](6.15,-0.18)--(6.15,-0.04);
	\draw[line width=0.35mm, blue](6.45,-0.34)--(6.45,-0.2);
	\draw[line width=0.35mm, blue](6.45,-0.18)--(6.45,-0.04);
	\draw[line width=0.35mm, blue](5.55,-0.34)--(5.55,-0.2);
	\draw[line width=0.35mm, blue](5.55,-0.18)--(5.55,-0.04);
	
	\draw[line width=0.35mm, blue](5.85,-0.42)--(5.85,-0.56);
	\draw[line width=0.35mm, blue](5.85,-0.58)--(5.85,-0.72);
	\draw[line width=0.35mm, blue](6.15,-0.42)--(6.15,-0.56);
	\draw[line width=0.35mm, blue](6.15,-0.58)--(6.15,-0.72);
	\draw[line width=0.35mm, blue](6.45,-0.42)--(6.45,-0.56);
	\draw[line width=0.35mm, blue](6.45,-0.58)--(6.45,-0.72);
	\draw[line width=0.35mm, blue](5.55,-0.42)--(5.55,-0.56);
	\draw[line width=0.35mm, blue](5.55,-0.58)--(5.55,-0.72);

\end{tikzpicture}

%% file: figures/horizontal_array.tex
\definecolor{aurometalsaurus}{rgb}{0.43, 0.5, 0.5}
\definecolor{pastelpink}{rgb}{1.0,  0.8353,  0.8353}

\begin{tikzpicture}
	\node[anchor=center] at (9,0) {\includegraphics[scale=0.3]{figures/ear_black.pdf}};

	\draw[line width=0.35mm, red](8.32,0.57)--(8.46,0.57);
	\draw[line width=0.35mm, red](8.48,0.57)--(8.62,0.57);
	\draw[red,line width=0.35mm](8.7,0.57)--(8.84,0.57);
	\draw[red,line width=0.35mm](8.86,0.57)--(9,0.57);
	\draw[line width=0.35mm, red](9.08,0.57)--(9.22,0.57);
	\draw[line width=0.35mm, red](9.24,0.57)--(9.38,0.57);
	\draw[line width=0.35mm, red](9.46,0.57)--(9.6,0.57);
	\draw[line width=0.35mm, red](9.62,0.57)--(9.76,0.57);
	
	\draw[line width=0.35mm, red](8.32,0.19)--(8.46,0.19);
	\draw[line width=0.35mm, red](8.48,0.19)--(8.62,0.19);
	\draw[line width=0.35mm, red](8.7,0.19)--(8.84,0.19);
	\draw[line width=0.35mm, red](8.86,0.19)--(9,0.19);
	\draw[line width=0.35mm, red](9.08,0.19)--(9.22,0.19);
	\draw[line width=0.35mm, red](9.24,0.19)--(9.38,0.19);
	\draw[line width=0.35mm, red](9.46,0.19)--(9.6,0.19);
	\draw[line width=0.35mm, red](9.62,0.19)--(9.76,0.19);
	
	\draw[line width=0.35mm, red](8.32,-0.19)--(8.46,-0.19);
	\draw[line width=0.35mm, red](8.48,-0.19)--(8.62,-0.19);
	\draw[line width=0.35mm, red](8.7,-0.19)--(8.84,-0.19);
	\draw[line width=0.35mm, red](8.86,-0.19)--(9,-0.19);
	\draw[line width=0.35mm, red](9.08,-0.19)--(9.22,-0.19);
	\draw[line width=0.35mm, red](9.24,-0.19)--(9.38,-0.19);
	\draw[line width=0.35mm, red](9.46,-0.19)--(9.6,-0.19);
	\draw[line width=0.35mm, red](9.62,-0.19)--(9.76,-0.19);
	
	\draw[line width=0.35mm, red](8.32,-0.57)--(8.46,-0.57);
	\draw[line width=0.35mm, red](8.48,-0.57)--(8.62,-0.57);
	\draw[line width=0.35mm, red](8.7,-0.57)--(8.84,-0.57);
	\draw[line width=0.35mm, red](8.86,-0.57)--(9,-0.57);
	\draw[line width=0.35mm, red](9.08,-0.57)--(9.22,-0.57);
	\draw[line width=0.35mm, red](9.24,-0.57)--(9.38,-0.57);
	\draw[line width=0.35mm, red](9.46,-0.57)--(9.6,-0.57);
	\draw[line width=0.35mm, red](9.62,-0.57)--(9.76,-0.57);
\end{tikzpicture}

%% file: data/APD_dipole.tex
	\definecolor{darkcandyapplered}{rgb}{0.64, 0.0, 0.0}
	\definecolor{darkpastelgreen}{rgb}{0.01, 0.75, 0.24}
	\definecolor{fluorescentpink}{rgb}{1.0, 0.08, 0.58}
	\definecolor{safetyorange(blazeorange)}{rgb}{1.0, 0.4, 0.0}
	\definecolor{blue-violet}{rgb}{0.54, 0.17, 0.89}
	\definecolor{darkbrown}{rgb}{0.4, 0.26, 0.13}
		\definecolor{bole}{rgb}{0.47, 0.27, 0.23}
\begin{tikzpicture}
     \begin{axis}[
	    xmin=2.5, xmax=17.5,
	    xlabel=Distance,
	    x unit=\si{\mm},
	    ymin=0, ymax=8,
   	    ylabel=$S_\text{ab}$,
	    y unit=\si{\W\per\square\m},
	    xtick={5,10,...,15},
	    unit markings=parenthesis,
	    tick label style={font=\boldmath},
	    legend style={at={(1,1)},anchor=north east},
	      y tick label style={
	    	/pgf/number format/.cd,
	    	fixed,
	    	fixed zerofill,
	    	precision=1
	    }
	    	    ]
	    
	    \addplot+[blue,only marks, mark=square, thick,mark size=3.5] table [x =distance, y expr=\thisrow{dipole_vertical}]{data/APD_dipole.txt};
		\addlegendentry{Vertical dipole}
		\addplot+[black, only marks, thick,  mark=o,mark options={fill=black},mark size=3.5] plot[error bars/.cd, y dir=both, y explicit]table [x =distance, y expr=\thisrow{array_vertical}]{data/APD_dipole.txt};
		\addlegendentry{Vertical array}

		\addplot+[darkpastelgreen, thick,mark size=3.5, only marks, mark=triangle*, mark options={solid,fill=darkpastelgreen}]table [x expr=\thisrow{distance}, y=dipole_horizontal]{data/APD_dipole.txt};
		\addlegendentry{Horizontal dipole} 
		
		\addplot+[red, thick, mark=square, mark size=4, only marks, mark=*, mark options={solid,fill=red}]  table [x expr=\thisrow{distance}, y=array_horizontal]{data/APD_dipole.txt};
		\addlegendentry{Horizontal array} 
		
	\end{axis}

\end{tikzpicture}